\documentclass{PoS}
\usepackage{amsmath,epsfig,indentfirst}
\newcommand{\eq}[1]{\begin{align} #1 \end{align}}

\title{Fluctuations and Correlations \\ from Microscopic Transport Theory}
\ShortTitle{Fluctuations and Correlations from Microscopic Transport Theory}

\author{\speaker{V.~P.~Konchakovski}\\
        Helmholtz Research School, University of Frankfurt, Germany\\
        Bogolyubov Institute for Theoretical Physics, Kiev, Ukraine\\
        E-mail: \email{voka@fias.uni-frankfurt.de}}

\author{M.~Hauer\\
        Helmholtz Research School, University of Frankfurt, Germany}

\author{M.~I.~Gorenstein\\
        Bogolyubov Institute for Theoretical Physics, Kiev, Ukraine\\
        Frankfurt Institute for Advanced Studies, University of Frankfurt, Germany}

\author{E.~L.~Bratkovskaya\\
        Institute for Theoretical Physics, University of Frankfurt, Germany}

\abstract{
The multiplicity fluctuations in A+A collisions at SPS and RHIC energies are studied within the HSD transport approach.
We find a dominant role of the fluctuations in the nucleon participant number for the final fluctuations.
In order to extract physical fluctuations one should decrease the fluctuations in the participants number.
This can be done considering very central collisions.
The system size dependence of the multiplicity fluctuations in central A+A collisions at the SPS energy range
-- obtained in the HSD and UrQMD transport models -- is presented.
The results can be used as a `background' for experimental measurements of fluctuations as a signal of the critical point.
Event-by-event fluctuations of the $K/\pi$, $K/p$ and $p/\pi$
ratios in A+A collisions are also studied.
Event-by-event fluctuations of the kaon to pion number ratio
in nucleus-nucleus collisions are studied for SPS and RHIC energies.
We find that the HSD model can qualitatively reproduce
the measured excitation function for the $K/\pi$ ratio fluctuations
in central Au+Au (or Pb+Pb) collisions from low SPS up to top RHIC energies.
The forward-backward correlation coefficient measured
by the STAR Collaboration in Au+Au collisions at RHIC is also studied.
We discuss the effects of initial collision geometry
and centrality bin definition on correlations in nucleus-nucleus collisions.
We argue that a study of the dependence of correlations
on the centrality bin definition as well as the bin size
may distinguish between these `trivial' correlations
and correlations arising from `new physics'.
}

\FullConference{
5th International Workshop on Critical Point and Onset of Deconfinement - CPOD 2009,\\
June 08 - 12 2009 \\
Brookhaven National Laboratory, Long Island, New York, USA
}

\begin{document}

\section{Introduction}

Nucleus-nucleus collisions at relativistic energies have been intensely studied over the last two decades.
The main goal of these efforts is to understand the properties
of strongly interacting matter under extreme conditions of high energy and baryon densities
for which the creation of a quark-gluon plasma (QGP) is expected~\cite{Co75,Sh80}.
%
%
Fluctuations of physical observables in heavy ion collisions (see e.g., the reviews \cite{He01,JeKo03})
may provide important signals regarding the formation of a QGP.
Measuring the fluctuations one might observe anomalies of the onset of deconfinement \cite{GaGoMr04,GoGaZo04}
and dynamical instabilities when the expanding system goes through the 1-st order transition line
between the quark-gluon plasma and the hadron gas \cite{JeKo03}.
Furthermore, the QCD critical point may be signaled by a characteristic pattern in fluctuations \cite{BaHe99}.
With the large number of particles produced in heavy ion collisions at CERN SPS and BNL RHIC energies
it has now become feasible to study fluctuations on an event-by-event basis.

Let's consider some region in phase space as an acceptance region for 
particles produced in the collision
and measure the multiplicity distribution within this acceptance.
Than to quantify multiplicity fluctuations one uses the scaled variance:
\eq{
\omega = \frac{\langle \left( N - \langle N \rangle \right)^2\rangle}{\langle N \rangle}
       = \frac{\langle N^2 \rangle - \langle N \rangle^2}{\langle N \rangle}~,
}
where $\langle \dots \rangle$ denoted event-by-event averaging.
By selecting particles  with certain quantum numbers
one can study fluctuations of electric charge, strangeness, charm, etc.

Ratio of two species (e.g $K/\pi$) in the acceptance also can be studied.
To characterize ratio fluctuations one usually uses different measures such as $\sigma$, $\nu$, $F$, $\Phi$, etc.
One can also analyze correlations between different species in terms of 
the correlation coefficient:
\eq{\label{rho}
\rho_{AB}~\equiv~\frac{\langle\Delta N_A~\Delta N_B\rangle}{\left[\langle\left(\Delta
N_A\right)^2\rangle~\langle\left(\Delta N_B\right)^2\rangle\right]^{1/2}}~.
}

Two different acceptance regions can be taken to study long range correlations.
For example forward-backward correlation measurements which have been 
performed by the STAR Collaboration and will be discussed here.
We note that higher moments of a distribution also can be used
to explore the early stage of the colliding system (see e.g. \cite{ScNaMiStBl09}).

A powerful tool for studying event-by-event physics is the transport model.
It allows to implement precisely an experimental acceptance,
study centrality and colliding energy dependences on final results, etc.
Results of two transport models will be shown here:
the Hadron-String-Dynamics (HSD) \cite{EhCa96,CaBr99} and
Ultra-Relativistic-Quantum-Molecular-Dynamics (UrQMD) \cite{Ba98,Bl99}.
These models provide a rather reliable description
(see, e.g. Refs. \cite{WeBrCaSt03,BrCaSt03,Ba98}) for the inclusive spectra of
charged hadrons in A+A collisions from SIS to RHIC energies.
Due to the fact that there is no explicit quark-gluon phase in HSD and UrQMD,
these models
can be used to examine the experimental data owith respect to the presence of `new physics'.
We mentioned that an explicit phase transition from hadronic to partonic degrees of freedom
is implemented in Parton-Hadron-String-Dynamics (PHSD)~\cite{CaBr09},
which will opens new possibilities in studying heavy-ion collisions.

\section{Multiplicity Fluctuations}

In each A+A collision only a fraction of all 2$A$ nucleons interact.
These are called participant nucleons and are denoted as
$N_P^{proj}$ and $N_P^{targ}$ for the projectile and target nuclei,
respectively. The nucleons, which do not interact, are called the
projectile and target spectators, $N_S^{proj} = A - N_P^{proj}$ and
$N_S^{targ} = A - N_P^{targ}$.

\begin{figure}[ht!]
\centerline{ \epsfig{file=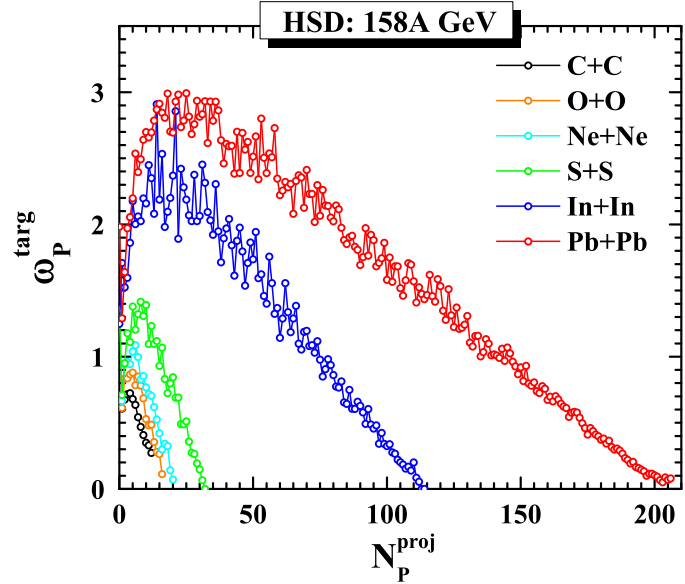,width=0.6\textwidth} }
\caption{The scaled variance $\omega_P^{targ}$  for the fluctuations of the number of target participants, $N_P^{targ}$. The HSD simulations of $\omega_P^{targ}$ as a function of $N_P^{proj}$ are shown for different colliding nuclei, In+In, S+S, Ne+Ne, O+O and C+C at $E_{lab}$=158~AGeV.}
\label{fig:w_targ}
\end{figure}

\begin{figure}[ht!]
\centerline{ \epsfig{file=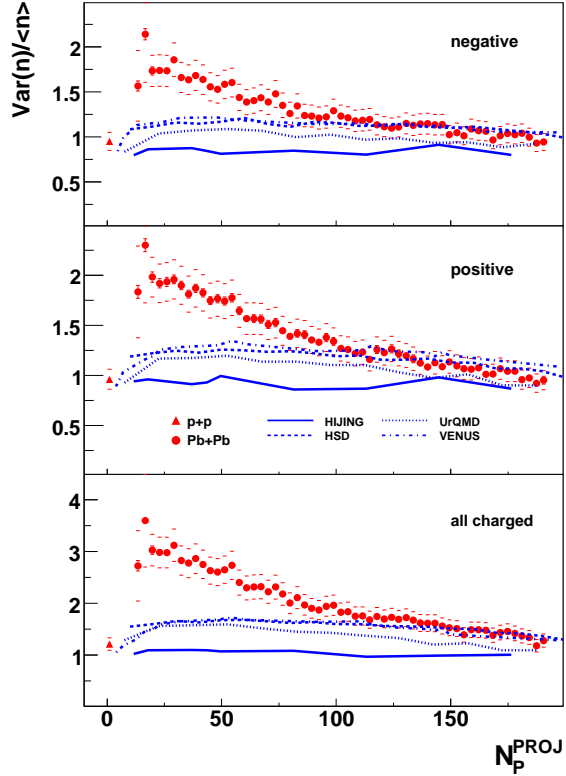,width=0.5\textwidth} }
\caption{The scaled variance of the
multiplicity distribution for negatively (upper panel), positively
(middle panel) and all (bottom panel) charged particles as a
function of the number of projectile participants $N_{P}^{PROJ}$
compared with model simulations in the NA49 acceptance (HSD and
UrQMD predictions were taken from~\cite{Ko06}). The
statistical errors are smaller than the symbols (except for the
most peripheral points). The figure is adapted from~\cite{Al07}.}
\label{fig:mult_NA49}
\end{figure}

In each sample with $N_P^{proj}=const$ the number of target participants fluctuates around its mean value,
$\langle N_P^{targ} \rangle$, with the scaled variance $\omega_P^{targ}$.
From an output of the HSD minimum bias simulations of A+A collisions at 158~AGeV
we form the samples of events with fixed values of $N_P^{proj}$.
Fig.~\ref{fig:w_targ} presents the HSD scaled variances $\omega_P^{targ}$ as functions of $N_P^{proj}$
for different colliding systems.
The fluctuations of target participants are quite strong for semi-peripheral 
collisions.
%
These large fluctuations in the number of participants strongly influence
the fluctuations of extensive observables such as multiplicity~\cite{Ko06},
electric charge and baryonic number~\cite{KoGoBrSt06}.
Experimental data of the NA49 Collaboration for the multiplicity fluctuations
also show an enhancement for semi-peripheral collisions (Fig.~\ref{fig:mult_NA49}).
Though both transport models HSD and UrQMD show a similar enhancement for 
the full acceptance,  this dependence become flat
when apply experimental acceptance (which is in the forward hemisphere). 
This effect can be explained by the nucleus-nucleus dynamics.


\begin{figure}[ht!]
\centerline{ \epsfig{file=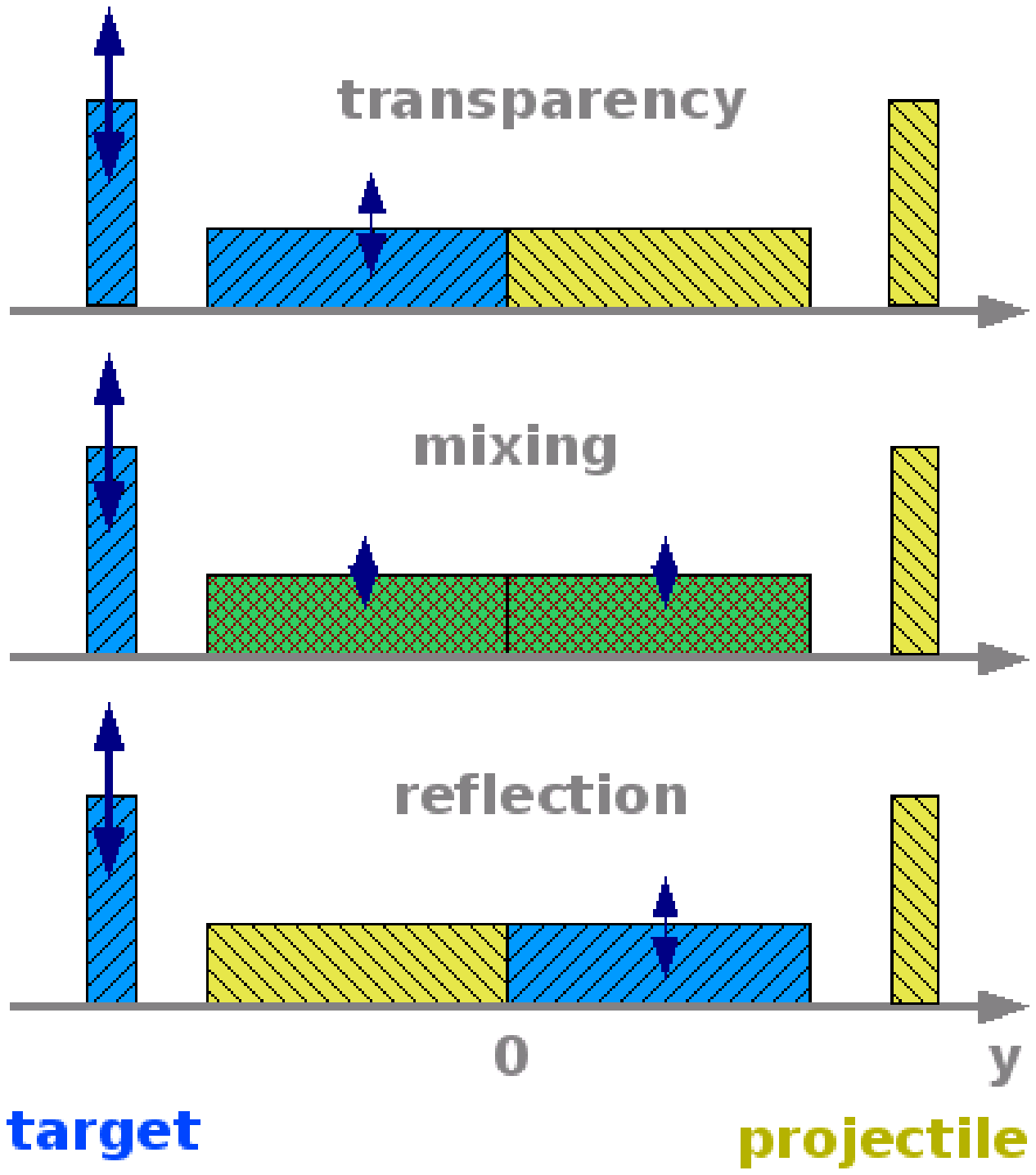,width=0.4\textwidth} }
\caption{The sketch of the rapidity distributions of the baryon number or the particle production sources (horizontal rectangles) in nucleus-nucleus collisions resulting from the transparency, mixing and reflection models. The spectator nucleons are indicated by the vertical rectangles. In the collisions with a fixed number of projectile spectators only matter related to the target shows significant fluctuations (vertical arrows). The figure is adapted from~\cite{GaGo06}.}
\label{fig:sketch}
\end{figure}

The consequences of the asymmetry between projectile and target hemispheres
-- introduced by fixing the number of projectile participants $N_P^{proj}$ -- depend on the A+A dynamics.
According to Ref.~\cite{GaGo06} different models of hadron production in
relativistic A+A collisions can be divided into three limiting
groups: transparency (T-), mixing (M-), and reflection (R-)
models. The rapidity distributions resulting from the T-, M-, and
R-models are sketched in Fig.~\ref{fig:sketch}.

The HSD model (as well as UrQMD) shows only a small mixing on initial baryon flow
and is closer to the T-model (cf. \cite{KoGoBrSt06}).
This supports the findings from Ref. \cite{WeBrCaSt03} about the influence of the partonic degrees of
freedom on the initial phase dynamics which should increase the mixing by additional strong parton-parton interactions.

To decrease fluctuations in the participant number (which result in observable fluctuations),
one needs to consider the most central collisions.
As seen from Fig.~\ref{fig:w_targ} for the most central collisions these fluctuations vanish.
Moreover one sees good agreement between data and transport models for more central collisions in Fig.~\ref{fig:mult_NA49}.


\begin{figure}[ht!]
\centerline{ \epsfig{file=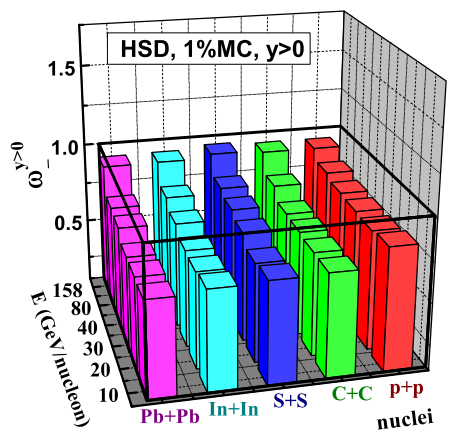,width=0.4\textwidth} }
\caption{
The HSD results for $\omega_-$ in $A+A$ and $p+p$ collisions for the rapidity $y>0$,
which is in correspondence with the NA61 program at SPS~\cite{Ga06}.
The 1\% most central collisions are selected
by choosing the largest values of $N_P^{proj}$ (see~\cite{KoLuGoBr08} for details).}
\label{fig:na61}
\end{figure}

An ambitious experimental program for the search of the QCD critical point
has been started by the NA61 Collaboration  at the SPS \cite{Ga06}.
The program includes collecting very central collisions varying in the 
atomic mass number $A$ and in the collision energy.
This allows to scan the phase diagram in the plane of temperature $T$
and baryon chemical potential $\mu_B$ near the critical point as argued in Ref.~\cite{Ga06}.
One expects to `locate' the position of the critical point by studying its `fluctuation signals'.
High statistics multiplicity fluctuation data will be taken for p+p,
C+C, S+S, In+In, and Pb+Pb collisions at bombarding energies of $E_{lab}$=10, 20, 30, 40, 80, and  158~AGeV.
We have considered these collision systems within the HSD (as well as UrQMD) transport model~\cite{KoLuGoBr08}.
The study thus is in full correspondence to the experimental program of the NA61 Collaboration \cite{Ga06}.
A monotonic energy dependence for the multiplicity fluctuations is obtained in the HSD transport model (Fig.~\ref{fig:na61}).
Thus, the expected enhanced fluctuations -- attributed to the critical point and phase transition --
can be observed experimentally on top of a monotonic and smooth `hadronic background'.
Our findings should be helpful for the optimal choice of collision
systems and collision energies for the experimental search of the QCD critical point.

\section{Ratio Fluctuations}

One of the possible measures to characterize fluctuations in the ratio $R_{AB}\equiv N_A/N_B$
is $\sigma$ \cite{BaHe99,JeKo99}:
\eq{\label{sigma-def}
\sigma^2~\equiv~\frac{\langle \left(\Delta R_{AB}\right)^2\rangle}
{\langle R_{AB}\rangle^2 } ~.
}
which can be expanded in the following way (see \cite{GoHaKoBr09} for details):
\eq{\label{sigma}
 \sigma^2~\cong~\frac{\omega_A}{\langle N_A\rangle}~+~\frac{\omega_B }{\langle
N_B\rangle}~-~2\rho_{AB}~\left[\frac{\omega_A \omega_B}{\langle
N_A\rangle\langle N_B\rangle}\right]^{1/2}~.
}

\begin{figure}[ht!]
\centerline{ \epsfig{file=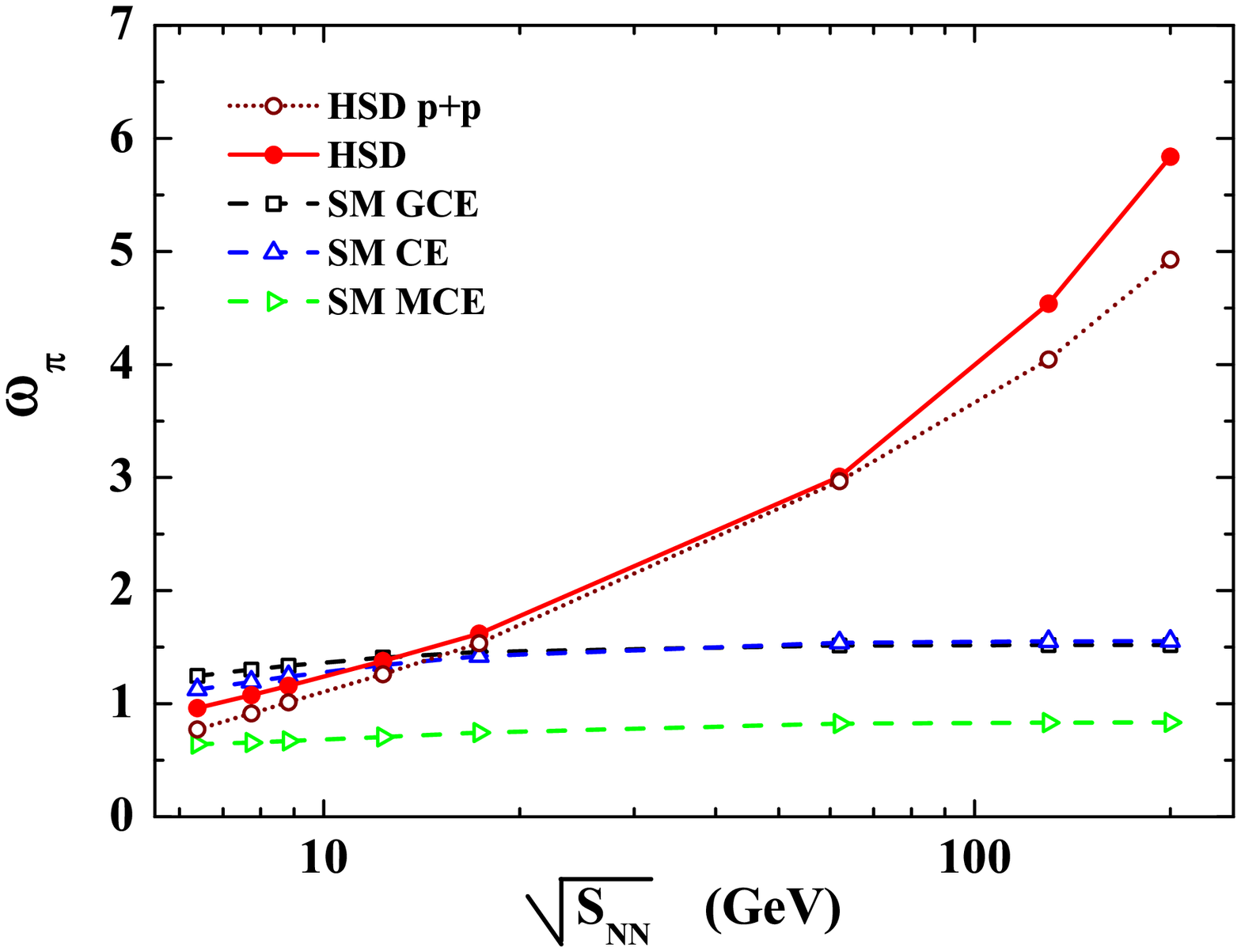,width=0.5\textwidth} \epsfig{file=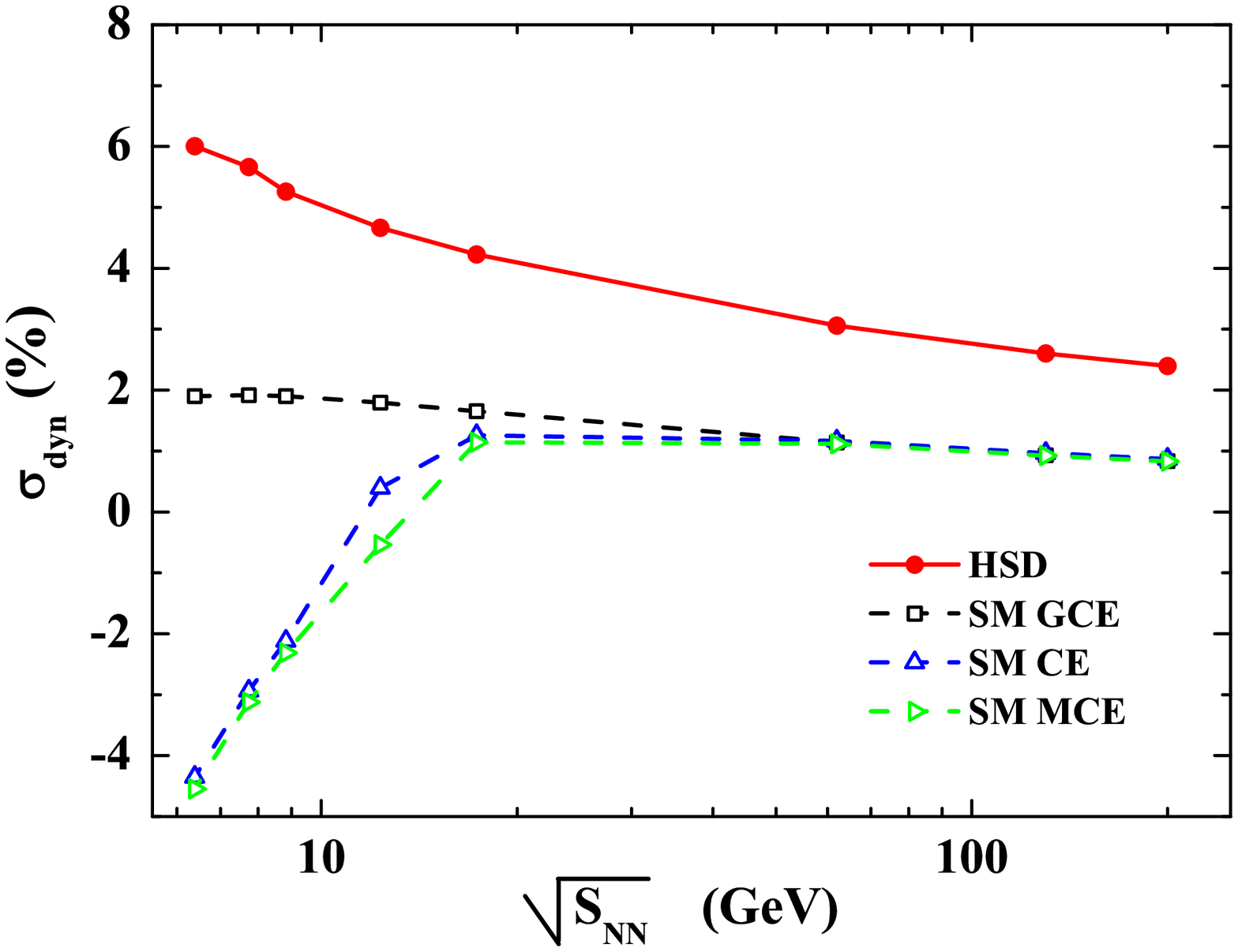,width=0.5\textwidth} }
\caption{The scaled variance $\omega_\pi$ of pions (left)
and $\sigma_{dyn}$ (\protect\ref{sigmadyn}) of $K/\pi$ fluctuations (right)
at different c.m. energies $\sqrt{s_{NN}}$
in the GCE, CE, and MCE ensembles (dashed lines) as well as from the HSD transport model (solid lines).}
\label{fig:omegaVSsigma}
\end{figure}

The experimental data for $N_A/N_B$ fluctuations are usually presented in terms of the so called
dynamical fluctuations~\cite{VoKoRi99}:
\eq{\label{sigmadyn}
\sigma_{dyn}~\equiv~\texttt{sign}\left(\sigma^2~-~\sigma^2_{mix}\right)\left|\sigma^2~-~\sigma^2_{mix}\right|^{1/2}~,
}
where $\sigma_{mix}$ corresponds to the {\it mixed events} procedure
which is calculated in the same way as $\sigma$~(\ref{sigma-def}) but for uncorrelated particles from different events.
Details about the mixed events procedure can be found e.g. in~\cite{GoHaKoBr09}.

\begin{figure}[ht!]
\centerline{ \epsfig{file=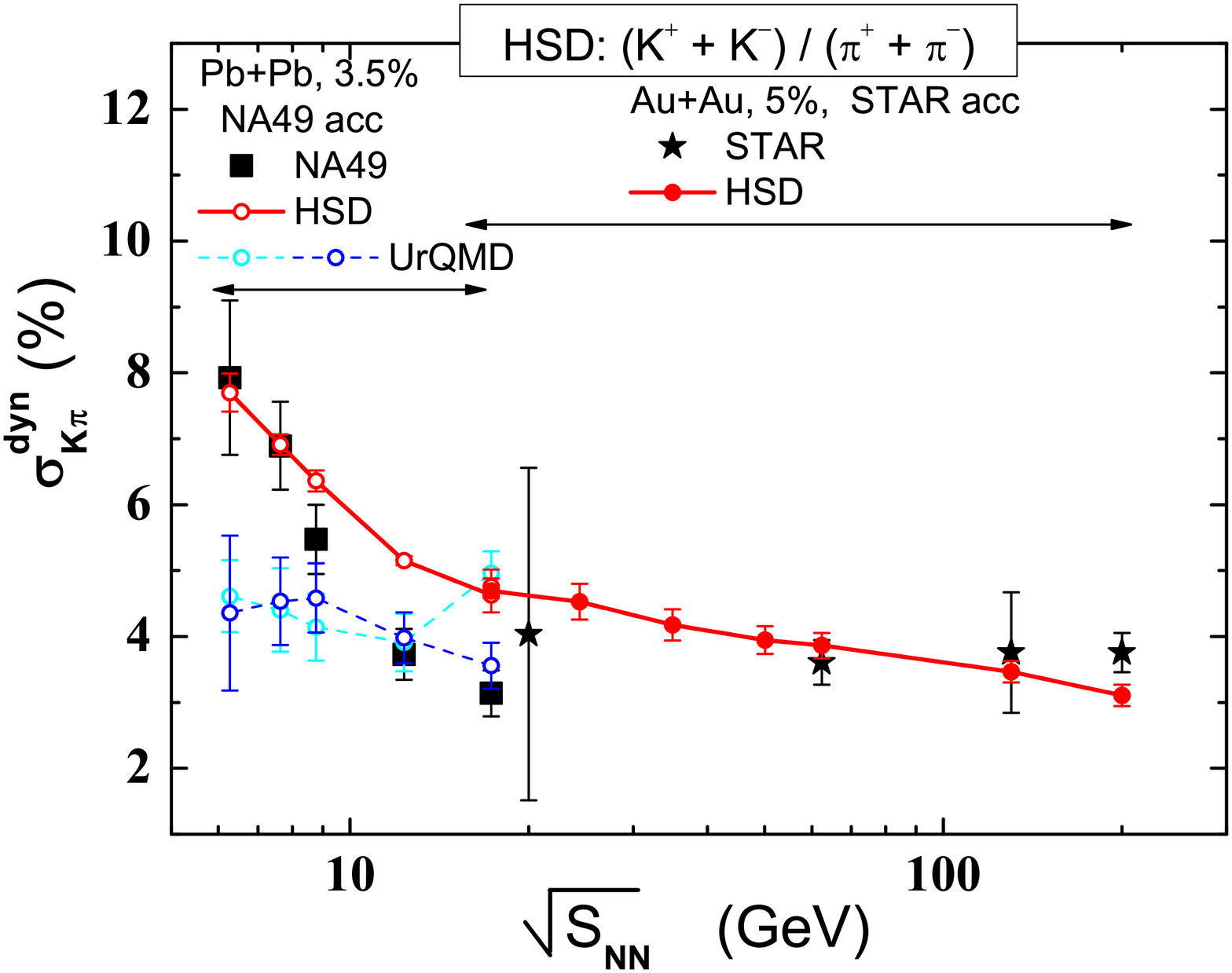,width=0.45\textwidth} }
\centerline{ \epsfig{file=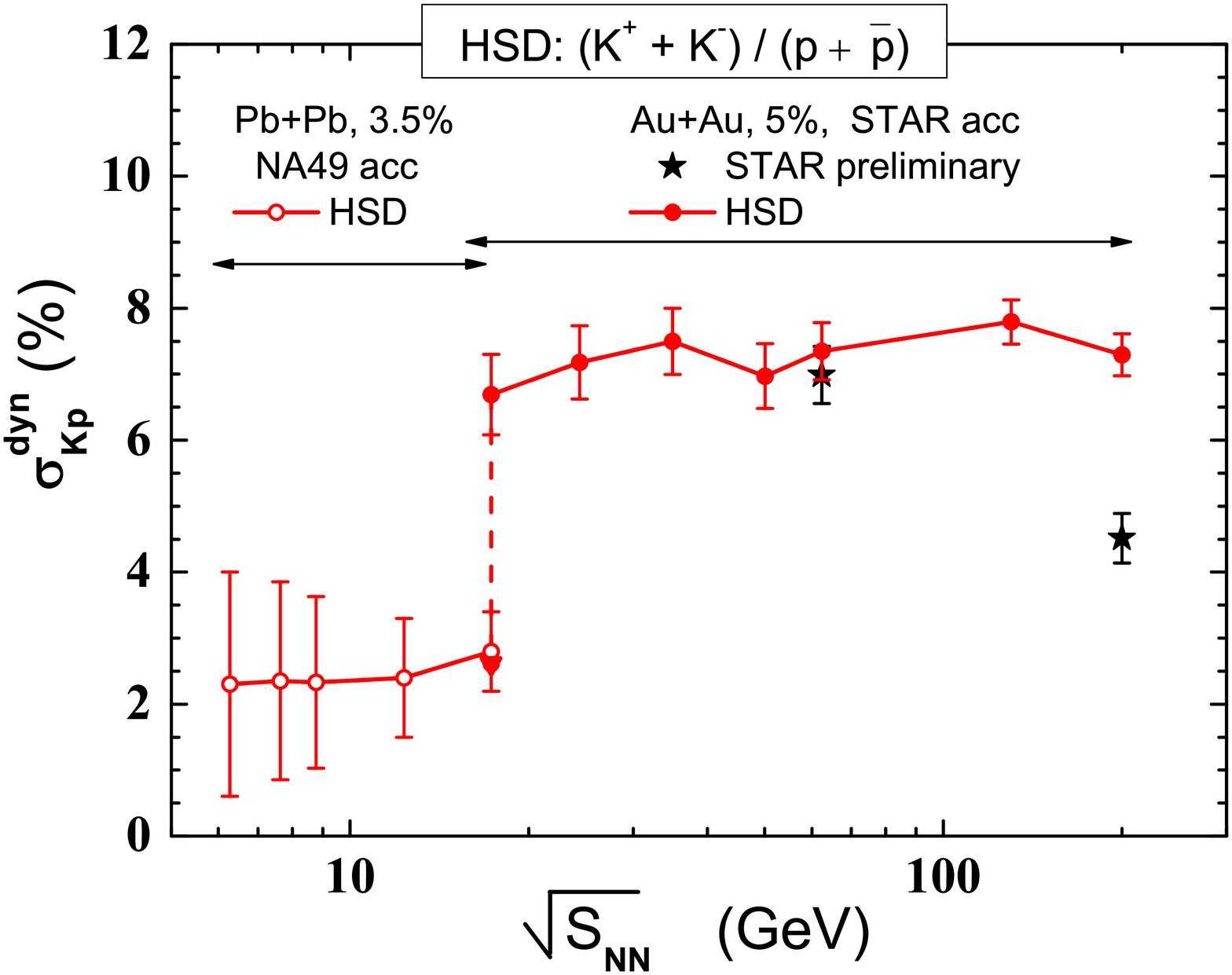, width=0.45\textwidth} }
\centerline{ \epsfig{file=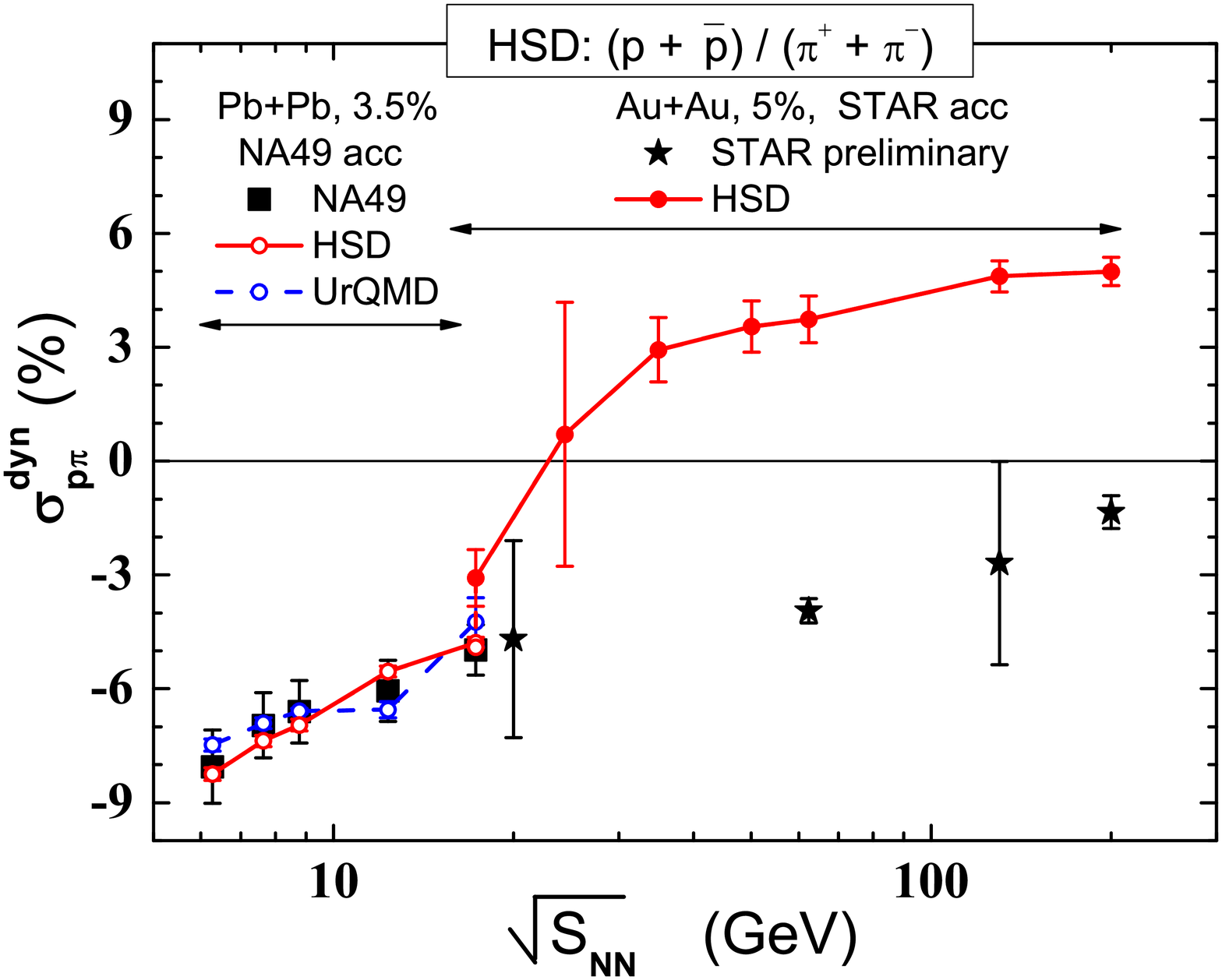,width=0.45\textwidth} }
\caption{The HSD results for the excitation function in $\sigma_{dyn}$ for the $K/\pi$, $K/p$, $p/\pi$ within the experimental acceptance (solid line) in comparison to the experimental data measured by the NA49 Collaboration at SPS~\cite{Al08} and by the STAR Collaboration at RHIC~\cite{Da06,Ab09,WeQM09,TiQM09}. The UrQMD calculations are shown by dotted lines. See \cite{GoHaKoBr09,KoHaGoBr09} for details.}
\label{fig:fig5}
\end{figure}

Fig.~\ref{fig:omegaVSsigma} presents the scaled variance $\omega_\pi$ of 
pions and $\sigma_{dyn}$ (\ref{sigmadyn}) of $K/\pi$ fluctuations.
Please note a different behavior of HSD model results for $\omega$ and 
$\sigma$ as a function of energy.
The scaled variance $\omega$ increases with energy while the dynamical fluctuations $\sigma$ decreases.
The reason is seen from Eq.~\ref{sigma}: $\sigma^2~\propto~\omega/\langle N \rangle$
where $\langle N \rangle$ is average multiplicity increasing with energy.
The difference between transport and statistical models also changes when we go from $\omega$ to $\sigma$.
The difference becomes larger with energy for the scaled variance~\cite{KoGoBr07}
but for $\sigma$ it is more pronounced for smaller energies.

In Fig.~\ref{fig:fig5} the HSD results of $\sigma_{dyn}$
for the $K/\pi$, $p/\pi$ and $K/p$ ratios are shown in comparison with
the experimental data by the NA49 Collaboration at the SPS~\cite{Al08}
and the preliminary data of the STAR Collaboration at RHIC~\cite{Da06,Ab09,WeQM09,TiQM09}.
The available results of UrQMD calculations (from Refs.~\cite{Al08,Ro04,KrFr06})
are also shown by the dashed lines.

The HSD results presented in Fig.~\ref{fig:fig5} correspond to the centrality
selection as in the experiment: the NA49 data correspond to the
3.5\% most central collisions selected via the veto calorimeter,
whereas in the STAR experiment the 5\% most central events with
the highest multiplicities in the pseudorapidity range
$|\eta|<0.5$ have been selected.

One sees that the UrQMD model gives practically a constant $\sigma_{dyn}^{K\pi}$,
which is by about $40\%$ smaller than the results from HSD at the lowest
SPS energy.
This difference between the two transport models may be attributed to
different realizations of the string and resonance dynamics in HSD and UrQMD: in
UrQMD the strings decay first to heavy baryonic and mesonic
resonances which only later on decay to `light' hadrons
such as kaons and pions.
In HSD the strings dominantly decay directly to
`light' hadrons  or the vector mesons $\rho$, $\omega$ and $K^*$.
Such a `non-equilibrated' string dynamics may lead to stronger fluctuations of
the $K/\pi$ ratio.

At the SPS energies the HSD simulations lead to negative values of $\sigma_{dyn}$ for the proton to pion ratio.
This is in agreement with the NA49 data in Pb+Pb collisions.
On the other hand HSD gives large positive values of $\sigma_{dyn}^{p\pi}$ at RHIC energies
which overestimate the preliminary STAR data for Au+Au collisions \cite{WeQM09}.
For $\sigma_{dyn}^{Kp}$ only preliminary STAR data in Au+Au collisions are available~\cite{TiQM09}
which demonstrate a qualitative agreement with the HSD results (Fig.~\ref{fig:fig5}).
The HSD results for $\sigma_{dyn}^{Kp}$ show a weak energy dependence in both SPS and RHIC energy regions.

An interesting feature is a strong `jump' between the SPS and RHIC values,
seen in the middle panel of Fig.~\ref{fig:fig5}, in the HSD calculations which is caused by the different acceptances in the SPS and RHIC measurements.
The influence of the experimental acceptance is clearly seen at
160 A GeV where a switch from the NA49 to the STAR acceptance leads
to the jump in $\sigma_{dyn}^{Kp}$ by 3\% - middle panel of Fig.~\ref{fig:fig5}.
On the other hand, our calculations
for Pb+Pb (3.5\% central) and for Au+Au (5\% central) collisions
- performed within the NA49 acceptance for  both cases at 160 A GeV -
shows only a very week sensitivity of $\sigma_{dyn}^{Kp}$ on the actual
choice of the collision system and centrality --
cf. the coincident open circle and triangle at 160 A GeV in the middle panel of Fig.~\ref{fig:fig5}.

\section{Forward-Backward Correlations}

Correlations of particles between different regions of rapidity
have for a long time been considered to be a signature of new physics.
The observation of such  correlations in A+A collisions at RHIC energies by the
STAR Collaboration \cite{Ta07,TaScSr07} has therefore elicited a lot of theoretical interest.

\begin{figure}[ht!]
\centerline{\epsfig{file=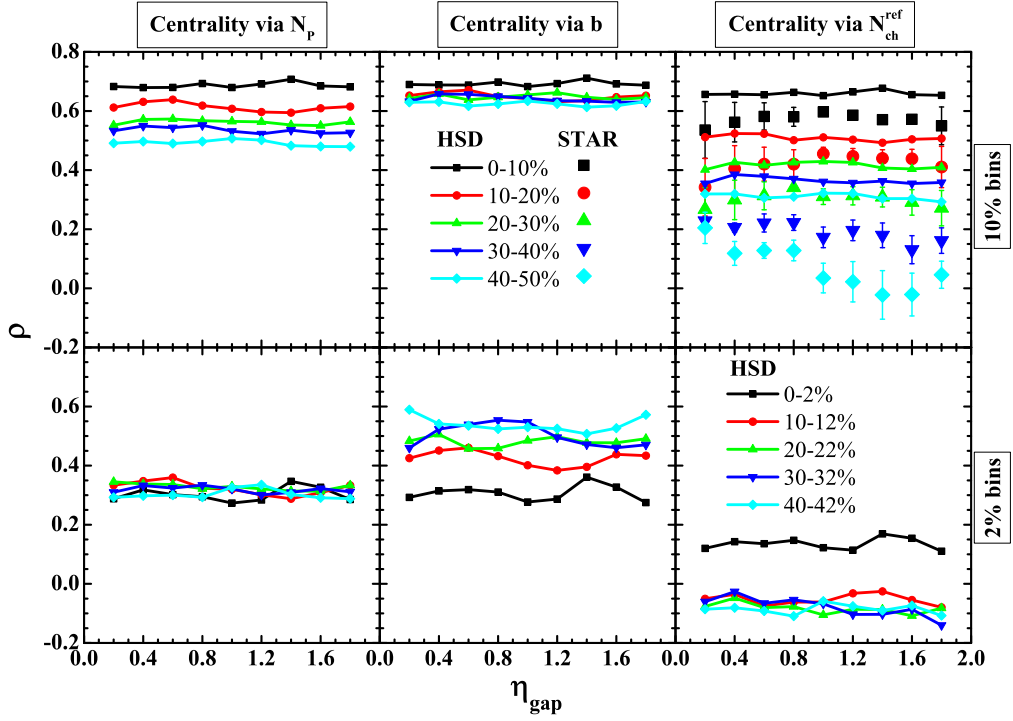,width=0.9\textwidth} }
\caption{The HSD results for the forward-backward correlation coefficient
$\rho$ for 10\% ({\it top}) and 2\% ({\it bottom}) centrality
classes defined via $N_{P}$ ({\it left}), via impact parameter $b$
({\it center}), and via the reference multiplicity $N_{ch}^{ref}$
({\it right}). The symbols in the {\it top right} panel present
the STAR data in Au+Au collisions at $\sqrt{s}=200$~GeV \cite{Ta07,TaScSr07}.}
\label{fig:HSD_FB_corr}
\end{figure}

\begin{figure}[ht!]
\centerline{ \epsfig{file=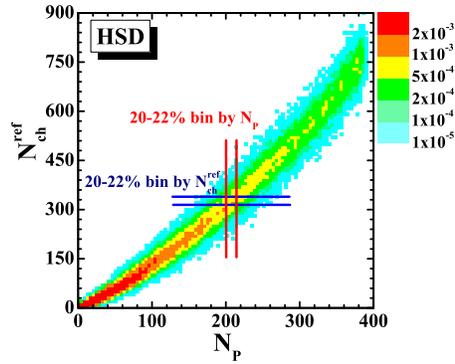,width=0.4\textwidth} }
\caption{The histogram shows the distribution of HSD events with fixed
number of participating nucleons $N_{P}$ and fixed reference
charge particle multiplicity $N_{ch}^{ref}$.  The same centrality
class (20-22\% as an example) - defined in various ways - contains
different events.}
\label{fig:HSD_Npart_Nref}
\end{figure}

Fig.~\ref{fig:HSD_FB_corr} shows the dependence of the
forward-backward correlation coefficient $\rho$ (\ref{rho}) as a function of
$\eta_{gap}$ on the bin size and centrality definition within the HSD model (see \cite{KoHaToGoBr08} for details).
The dependence of $\rho$ on $\eta_{gap}$ is almost
flat, reflecting a boost-invariant distribution of particles
created by string breaking in HSD.
The {\it right top} panel of Fig.~\ref{fig:HSD_FB_corr} demonstrates
also a comparison of the HSD results with the STAR data
\cite{Ta07,TaScSr07}. One observes that the
HSD results exceed systematically the STAR data. However, the main
qualitative features of the STAR data -- an approximate
independence of the width of the pseudo-rapidity gap $\eta_{gap}$
and a strong increase of $\rho$  with centrality -- are fully
reproduced by the HSD simulations.
Note that choosing smaller centrality bins leads to weaker forward-backward
correlations, a less pronounced centrality dependence, and a
stronger dependence on the bin definition.
The physical origin for this is demonstrated in
Fig.~\ref{fig:HSD_Npart_Nref}. As the bin size becomes comparable to
the width of the correlation band between $N_{P}$ and
$N_{ch}^{ref}$, the systematic deviations of different centrality
selections  become dominant: the same centrality bins defined by
$N_P$ and by $N_{ch}^{ref}$ contain different events and may give
rather different values for the forward-backward correlation
coefficient $\rho$.

Thus, the experimental analysis for different bin sizes and
centrality definitions -- as performed here -- may serve as a
diagnostic tool for an origin of the observed correlations. A
strong specific dependence of the correlations on  bin size and
centrality definition would signify their geometrical origin!

\section{Conclusions}

Our analysis has shown that the fluctuations in the number of participants strongly influence
observed multiplicity fluctuations. To avoid them one should consider the most
central collisions with rigid events selection.

We have considered C+C, S+S, In+In, and Pb+Pb nuclear collisions from $E_{lab}$= 10, 20, 30, 40, 80, 158~AGeV,
which is in full correspondence with the experimental program SHINE started at SPS.
A monotonic energy dependence for the multiplicity fluctuations
is obtained in the HSD transport model.
Thus, the expected enhanced fluctuations -
attributed to the critical point and phase transition - can be
observed experimentally on top of a monotonic and smooth `hadronic background'.
Our findings should be helpful for the optimal choice of collision
systems and collision energies for the experimental search of the QCD critical point.

It has been found that the HSD model can qualitatively reproduce the measured
excitation function for the $K/\pi$ ratio fluctuations in central A+A collisions
from low SPS up to top RHIC energies.
Accounting for the experimental acceptance as well as the
centrality selection has a relatively small influence on $\sigma_{dyn}$
and does not change the shape of the $\sigma_{dyn}$ excitation function.
The HSD results for $\sigma_{dyn}^{p\pi}$ also appear to be close to the NA49 data at the SPS.
On the other hand a comparison of the HSD results with preliminary STAR data in Au+Au
collisions at RHIC energies is not fully conclusive:
$\sigma_{dyn}$ from HSD calculations is approximately in
agreement with data \cite{TiQM09} for the kaon to proton ratio, but
overestimates the experimental results \cite{WeQM09} for the proton
to pion ratio. New data on event-by-event fluctuations in Au+Au at
RHIC energies will help to clarify the situation.

The forward-backward correlations have been studied within the HSD transport model for $\sqrt{s}=200~GeV$.
It has been shown that strong forward-backward correlations arise due to an averaging
over many different events that belong to one 10\% centrality bin.
In contrast to average multiplicities, the resulting fluctuations
and correlations depend strongly on the specific centrality
trigger.
When the size of the bins decreases, the contribution of `geometrical' 
fluctuations should lead to weaker forward-backward correlations and to 
a less pronounced centrality dependence. Note, that the `geo\-metrical' 
fluctuations discussed here are in fact present in all dynamical models 
of nucleus-nucleus collisions.  Thus, they should be carefully 
accounted for before any discussion of new physical phenomena is addressed.

%

\end{document}